# Lifestyle Pattern Analysis Unveils Recovery Trajectories of Communities Impacted by Disasters

**Natalie Coleman\*, Chenyue Liu, Yiqing Zhao, Ali Mostafavi**

[1] Ph.D. Student, Zachry Department of Civil and Environmental Engineering, Urban Resilience.AI Lab, Texas A&M University, College Station, TX, United States of America; email: ncoleman@tamu.edu
\*Corresponding author
[2] Ph.D. Student, Zachry Department of Civil and Environmental Engineering, Urban Resilience.AI Lab, Texas A&M University, College Station, TX, United States of America; email: liuchenyue@tamu.edu
[3] Masters, Department of Computer Science and Engineering, Engineering, Urban Resilience.AI Lab, Texas A&M University, College Station, TX, United States of America; email: yiqingzhao@tamu.edu
[4] Associate Professor, Zachry Department of Civil and Environmental Engineering, Urban Resilience.AI Lab, Texas A&M University, College Station, TX, United States of America; e-mail: amostafavi@civil.tamu.edu

**ABSTRACT**

The return of normalcy to the population's lifestyle is a critical recovery milestone in the aftermath of disasters, and delayed lifestyle recovery could lead to significant well-being impacts. Lifestyle recovery captures the collective effects of population activities and the restoration of infrastructure and business services. This study uses a novel approach to leverage privacy-enhanced location intelligence data, which is anonymized and aggregated, to characterize distinctive lifestyle patterns and to unveil recovery trajectories after a disaster in the context of the 2017 Hurricane Harvey in Harris County, Texas (USA). The analysis integrates multiple data sources to record the number of visits from home census block groups (CBGs) to different points of interest in the county during the baseline period (first two weeks of August 2017) and after the disruptive period (following the fourth week of August 2017). First, primary clustering using k-means characterized four distinct essential and non-essential lifestyle patterns. For each primary lifestyle cluster, the secondary clustering characterized the impact of the hurricane into three recovery trajectories based on the severity of maximum disruption and duration of recovery. In addition, the analysis calculated the statistical significance of the results through ANOVA means testing. The results show distinctive lifestyle characterizations in Harris County. They further reveal multiple recovery trajectories and durations within each lifestyle cluster, which imply differential recovery rates among similar lifestyles and different demographic groups. The findings also show that the impact of flooding on lifestyle recovery extends beyond the flooded regions, as 59% of CBGs with long recovery periods (>15 weeks) did not have at least 1% of direct flooding impacts, meaning that at least 1% of the land area of the CBG experienced flooding impacts. The findings offer a twofold theoretical significance: (1) lifestyle recovery is a critical milestone that needs to be examined, quantified, and monitored in the aftermath of disasters; (2) the spatial structures of cities formed by human mobility and distribution of facilities extends the spatial reach of flood impacts on population lifestyles. The analysis and findings also provide novel data-driven insights for public officials and emergency managers to examine, measure, and monitor a critical milestone in community recovery trajectory based on the return of lifestyles to normalcy.

**INTRODUCTION**

Disasters cause not only physical damage to infrastructure systems and building facilities but also affect the societal dynamics of communities by disrupting the lifestyles of populations. Lifestyles can be characterized by essential and non-essential activities, such as visiting grocery stores to meet food needs or visiting local shopping centers to buy clothing, respectively. The disruption of essential lifestyles negatively impacts populations, as they are unable to access critical resources to maintain their health and well-being. The restoration of non-essential services can also signal a return to normalcy, as people are more able and comfortable to resume their previous activities. During the rebuilding and recovery process following a disaster, people may incrementally resume their standard lifestyles. The return to pre-disruption lifestyle standards could represent an important facet of the societal impacts of disasters, and thus, has the potential to unveil a critical milestone in community recovery [1,2].

Recovery is a vital component of the disaster management cycle. According to FEMA [3], disaster recovery involves the return of all formal and informal systems to a normal state. Eid and El-adaway [4] show disaster recovery is multidimensional and nonlinear, involving strategies and capacities to rebuild community livelihoods. Indeed, it is well-established that recovery is a complex process consisting of multiple dimensions [5-7, 10]. In addition, previous recovery frameworks have singled out the importance of recognizing people's life activities as an important element for measuring recovery as a community rebuilds after a disaster.[7, 8] Finucane [9] discusses how recovery planning should be adaptive to such needs and should establish formal mechanisms for assessment. However, there is a gap of knowledge related to the specification, characterization, and quantification of recovery milestones to facilitate proactive monitoring. The analysis of lifestyle patterns before and after disasters would bring novel and unique insights about the time at which the lifestyle patterns of impacted a population's returns to normalcy. Due to their holistic and dynamic nature, lifestyles encompass routine changes across various spatial units after the disaster impact. In particular, lifestyle recovery represents a critical milestone based on capturing people's interactions with infrastructure, businesses, and social interactions in the community [10,11]

Location intelligence data can reveal important insights about complex disaster phenomena pertaining to population activities and interactions with infrastructure[1,12,13]. For example, location intelligence data has been utilized to examine disaster impacts [14-16], urban flooding, and damage reporting[17,18], evacuation patterns of populations [19-21], and rapid impact assessments[22]. For instance, Hong [21] categorized the evacuation patterns caused by Hurricane Harvey into shelter-in-place, stable, distressed, and abandoned households. Indeed, certain research studies found that human mobility has distinctive patterns during stable and disruptive periods[23,24]. In addition, Podesta [10] measured periods of disruption to points of interest (POIs) in Harris County at the county level after the landfall of Hurricane Harvey; the study concluded that POI systems have differential recovery rates to their baselines. To expand on the understanding of accessibility to essential services, Esmalian et al. [25] and Fan et al. [12], utilizing location intelligence data to determine the equitable accessibility to critical facilities during disruptive events, found unique time-series clusters of the examined users. Despite these recent advances, little of the existing work has leveraged location intelligence data for specifying and characterizing community recovery. Unlike traditional datasets (e.g., surveys) used for examining recovery, location intelligence data has greater granularity, larger samples, shorter lags in data collection; and the use of this data could relive the burden of data collection from impacted populations. Also, location intelligence data could reveal important critical recovery milestones[26]. In particular, examining lifestyle recovery as a critical recovery milestone, can provide important insights about recovery trajectories at a granular level.

In this study, we leveraged location intelligence data, specifically POI data and human mobility data, to examine lifestyle activities at the census block group (CBG) level. In doing so, this study presents a novel methodology to specify lifestyle patterns and understand lifestyle recovery as a critical milestone following a disaster. We processed and analyzed datasets from Spectus and SafeGraph, two prominent commercial companies that collect and process location-based data at a granular scale. We combined that data with publicly available building footprint data to specify population visitations to POI and to characterize lifestyle patterns in the context of the Hurricane Harvey in Harris County, Texas, from August 2017 through November 2017. The analysis focused on answering the following research questions: (1) What are distinctive patterns in typical lifestyles of populations during normal periods?; (2) To what extent do disasters impact lifestyle activities and how quickly do lifestyle patterns recover?; and (3) To what extent do the duration and trajectory of lifestyle recovery vary across different areas with different flooding impact and demographic attributes?

**DATA AND METHODS**
**Study context**
We analyzed and quantified the recovery of lifestyles through the use of location intelligence data captured before, during, and after the 2017 Hurricane Harvey in Harris County, Texas. Hurricane Harvey was a Category 4 storm that made landfall in August 2017 and greatly impacted the Texas coastal region. The extensive flooding led to inaccessibility of essential and nonessential facilities[27]. The Texas Department of Transportation reported nearly 350 Houston area[28,29] road locations affected, and ERCOT reported thousands of power disruptions[30]. These disruptions had significant impacts on people's lifestyle activities in the aftermath of Harvey. The extensive impacts of Harvey and the metropolitan setting of Harris County provide a suitable context for examining lifestyle recovery patterns[31].

**Data sources**

To specify daily lifestyle patterns, we processed multiple datasets to determine daily visitations from each CBG to different POIs. Figure 1 depicts an overview of the methods for processing and analyzing the datasets. Human mobility data was primarily obtained from Spectus[32], a location intelligence data company which provided location data (e.g., GPS data) of high spatial accuracy and very high frequency of observations. Every day, more than 100 data points on average are collected from each anonymous device. Currently, there are roughly 15 million daily active users in the United States. The Spectus panel consists of users who provided informed consent to anonymized data collection for research purposes, through a GDPR and CCPA compliant framework. Compared to traditional Call Detail Record (CDR) data, which provides general information about telephone exchanges and interactions, the location intelligence data with precise GPS information provides exact data about destinations, which is extremely useful for detecting human lifestyle patterns in detail. Spectus's main database is built using third-party apps that capture opted-in users' anonymized location points if they consent to share their location information with these apps. Data is anonymized and aggregated at the CBG level to ensure consumer privacy and confidentiality.

For 2017, Spectus data lacked North American Industry Classification System (NAICS) codes to distinguish the type of each POI (stop points of unique device IDs). To address this limitation, we utilized the location and type of different POIs consistent with NAICS codes from the SafeGraph dataset [33]. Safegraph's core dataset contains spatial coordinates and addresses of POIs, as well as basic information about each POI, such as brand, operating time, and NAICS code. SafeGraph gathers POIs by scraping open store locators on the web using publicly available APIs and scraping open web domains that provide updated locations. To fill in the gaps, SafeGraph independently performs processing and modeling as well as licenses third-party data to infer additional attributes. After collecting the POIs from Safegraph's core dataset, we used Microsoft's open building footprint datasets [34] to determine the polygon of each POI. The 1,542,887 building footprint polygon geometries in the Houston metropolitan area were created by Microsoft using their computer vision algorithms on satellite imagery.

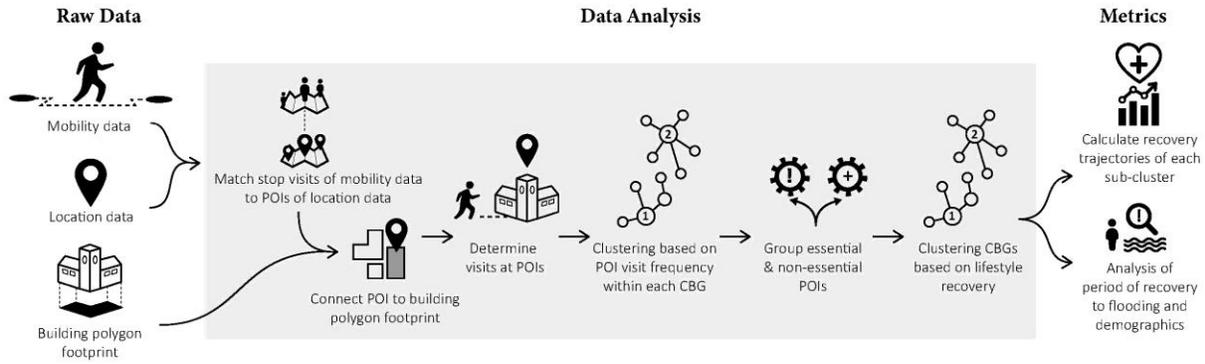

**Figure 1.** Integrating multiple data sources to specify population visitation to POI for characterizing lifestyle patterns
**Notes:** Raw data includes the multiple data sources to specify population visitation to POIs for characterizing lifestyle patterns. Raw data includes the collected multiple data sources. Data analysis includes merging of data sources for visits at POIs and the overall methods for clustering the data. Metrics are the final outputs of the methods including calculating the recovery trajectories.

For evaluating the spatial variation of lifestyle impacts and recovery based on flooding status, we used flooding data based on the estimated flood depths on August 29, 2017, from FEMA open-source data[35]. The data had a gridded horizontal resolution of 3 meters (m), which was processed appropriately to the analysis at CBG scale. We used this data to determine the percentage of CBG area flooded during Harvey. This percentage served as an indicator of flooding extent in each CBG. Demographic information, including total population, median income, percentage of white population, and percentage of elderly residents above 65 years old, were collected at a CBG level from the US Census data[36].

**Data processing of human visitation to POIs**
*Home_CBG to stop table:* Data processing focused on the period of August to November 2017. The cut-off date was the end of November to eliminate the effect of the holiday season on population lifestyle patterns. To identify each anonymous device's home CBG in Harris County, we first retrieved data from the device matrix table in the Spectus core database. We recognized a CBG as a user's home_CBG if the device consistently resided there for more than one day. Device_id and home_CBG id are included in the device matrix. Second, we used the Spectus core dataset's stop table to capture anonymous human mobility. The stop points were captured in the dataset if individuals stop at a location period for an appropriate time scale. Device_id, latitude and longitude, stop-by date, and stop-by time were all stored in the stop database. By merging the device matrix table and stop table through the device_id, home_CBG was acquired.
*POI polygon table:* To specify the polygon for each POI, the place_id in Spectus's dataset was extracted using the POI table in Spectus's core datasets. This table contains basic information about each POI, such as place id, place name, place address, and location's latitude and longitude. We obtained each place id with its associated building polygon by merging the POI table with the building footprint dataset.

*Home_CBG to POI table:* By merging the latitude and longitude of the POI table to the building footprint dataset, we matched each stop point of POI with a unique place_id. Then we classified each POI using the NAICS code [37], which provides information on the associated sector, subsector, industry group, and industries. Since the 2017 Spectus data lacks the NAICS code, we can match the polygon information (latitude and longitude) to Safegraph's data, which does include NAICS code. Informed by prior studies[10,38], the study identified 11 POIs which contained 5 essential services (gasoline stations, grocery and merchandise, health and personal care stores, medical facilities, and education) and 6 non-essential services (banks, stores and dealers, restaurants, entertainment, recreation and gym centers and beauty care). The detailed POI-NAICS table can be found at Table SI.1 in Supplementary Information. By doing so, the daily visits data from the home_CBG to POI was generated, and weekly data in this phase was generated by aggregating 7 days of daily data. The rationale for weekly aggregation is that lifestyle patterns might vary between weekdays but is usually consistent on weekly basis[38,39].

To analyze and quantify lifestyle recovery for different CBGs, we first established a baseline against which the lifestyle impacts could be compared. The baseline was calculated by averaging the number of visits during the first two weeks of August, the period before Hurricane Harvey made landfall. Second, we grouped POIs into essential and non-essential POIs to establish the baseline number of visits to each group of POIs. Then, we compared the number of visits in the subsequent weeks against the baseline to measure lifestyle recovery. A CBG is considered recovered if the number of visits to POIs is 90% of the baseline value.

**Primary and secondary k-means clustering**
Instead of analyzing the lifestyle patterns of each CBG separately, we first used k-means clustering for categorizing CBGs based on the frequency of visits to essential and non-essential POIs to establish lifestyle patterns. K-means clustering is an unsupervised learning method for determining clusters and cluster centers in a set of unlabeled data [12,40-43].

As shown in Figure 2, we used k-means clustering in two steps. The first step determined the clusters of CBGs based on their non-disrupted lifestyle patterns before Harvey; the second step clustered the CBG within each primary cluster based on the duration of lifestyle recovery after Harvey. To minimize the impact of outliers, the visitation values were standardized according to a minimum-maximum standardization formula. The primary clustering was obtained based on the POI visitations during the baseline period of the first two weeks of August 2017 (before the landfall of Hurricane Harvey). Vector data consisted of the aggregate frequency of visits to POIs over the total number of visits to all POIs. For example, the first input in the vector would consider the number of visits to grocery stores divided by the total number of visits to all POIs for a particular CBG within the same time period. The secondary clustering of the lifestyle recovery is based on four features: maximum point of disruption in essential POI visits, maximum point of disruption in non-essential POI visits, duration of recovery in essential POI visits, and duration of recovery in non-essential POI visits. Maximum point of disruption is defined as the greatest percent change in visits to essential and non-essential POIs and was recovered for the three weeks of . Duration of recovery is the number of weeks that for CBGs to recover to 90% of the baseline value. In both clustering methods, the number of clusters is determined through the elbow method as shown in Figure SI.1 in the Supplementary Information.

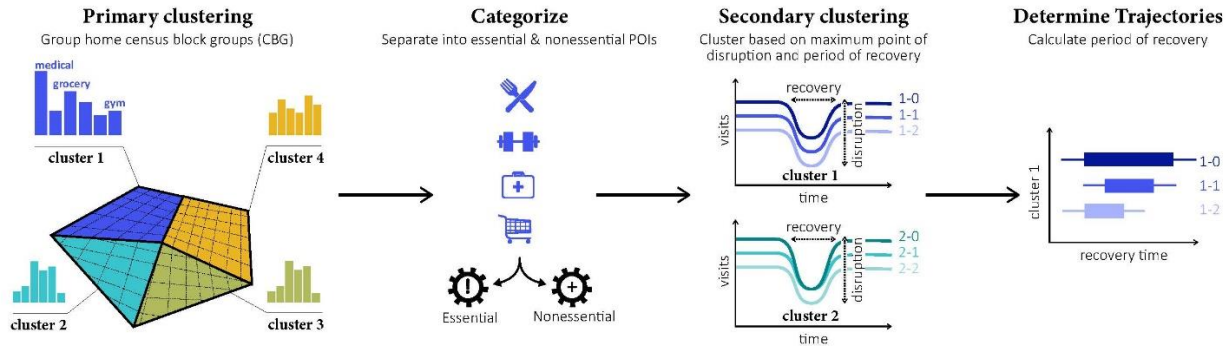

**Figure 2.** Primary and secondary clustering to determine the extent of lifestyle characterization and disruption
**Notes:** Primary clustering groups home CBGs are based on prominent visits to POIs. Then the POIs are grouped into essential and non-essential lifestyles. Secondary clusters groups within each of the primary clusters based on the maximum point of disruption and duration of recovery for essential and non-essential POIs. Finally, trajectories of recovery are examined based on mean and standard deviation.

**RESULTS**
Primary clustering of k-means can further characterize the lifestyles of CBGs for the baseline period before Hurricane Harvey. The elbow method (Figure SI.1 in Supplementary Information) determined that four clusters were the optimal number of k-clusters. The median values and distribution of the demographic characteristics of the four clusters are found at Figure SI.2 and Table SI.2 in  Supplementary Information. In essence, 60.26% of CBGs were in cluster 1; 3.92% in cluster 2; 27.94% in cluster 3, and 8.11% in cluster 4. According to the merged census data, Cluster 1 contained about 3.09 million people within an area of $2.86*10^9$ square miles; Cluster 2 contained $1.42*10^5$ people within an area of $6.46*10^7$ square miles; Cluster 3 contained about 1.48 million people within an area of 2.87 with $1.48*10^9$ square miles; and Cluster 4 contained about $3.21*10^5$ people within an area of $2.05*10^8$ square miles. Figure 3 shows the spatial distribution and relative frequency of visits of the essential and non-essential lifestyles. The exact values of the visit frequency to the essential and non-essential POI categories are tabulated in the Supplementary Information for Tables SI.3a and SI.3b. Non-essential categories had consistent rankings across the four clusters with the top three rankings as stores and dealers, restaurants, and beauty care. The essential categories had consistent rankings for Cluster 1, 3, and 4 with the top three being health and personal care stores, grocery and merchandise, and medical facilities. However, Cluster 2 had a slight adjustment with the top three ordered rankings as health and personal care stores, medical facilities, and grocery and merchandise .

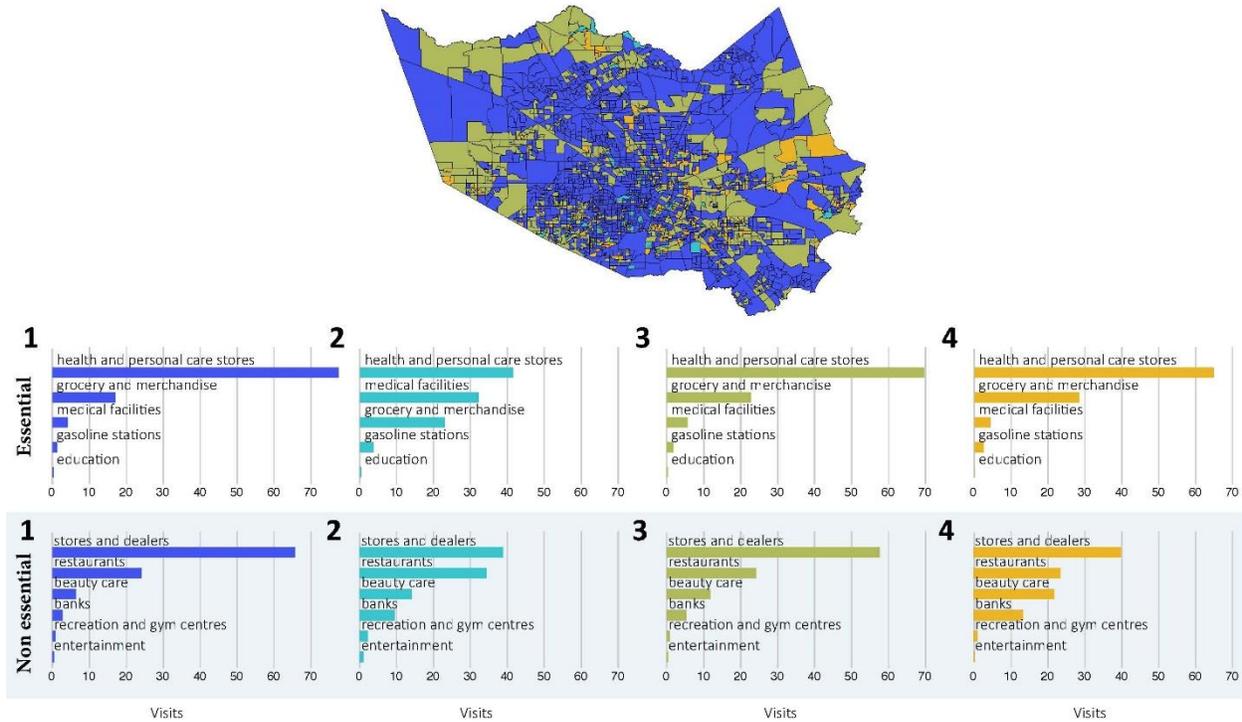

**Figure 3.** Results show the frequency distribution of the essential and non-essential lifestyles across the four primary clusters

**Notes:** The POIs with greater visits those upon whose services the populations are more dependent. The bar plots (a-h) present the prominent weekly POI visitations (essential and non-essential) within each lifestyle cluster. The X-axis shows the number of weekly visitation and y-axis shows the type of POI.

Table 1 displays how the relative frequency of each lifestyle varied across the clusters, which was confirmed by analysis of variance (ANOVA) testing a statistical p-value of < 0.05. This indicates that each primary cluster has distinct dependence on its lifestyles when compared to the other clusters. It is important to recognize the subtle differences between the primary clusters that make up the essential and non-essential lifestyles. Cluster 1 had no distinct high frequency similarity between the essential and non-essential lifestyles. When compared to other clusters, Cluster 2 had the highest relative frequency for essential services of medical facilities (57%) and self-care (19%) as well as for non-essential services of restaurant (27%) and entertainment (0.60%). Cluster 3 had the highest relative frequency for essential services of health and personal care stores (77.40%) and education (.32%) as well as for non-essential services of stores and dealers (65.57%). When compared to other clusters, Cluster 2 had the highest relative frequency for essential services of medical facilities (31.97%) and gasoline stations (3.60%) as well as for non-essential services of restaurant (34.27%), recreation and gyms (2.21%), and entertainment (0.99%). Cluster 3 had no distinct high frequency to the essential and non-essential lifestyles. Lastly, Cluster 4 had the highest relative frequency for essential services of grocery and merchandise (28.40%) and nonessential services of beauty care (21.86%) and banks (13.37%).

**Table 1.** ANOVA testing on frequency of lifestyles for primary clusters

| Essential POIs | Essential F-value | Non-essential POIs | Non-Essential F-value |
| --- | --- | --- | --- |
| Education | 32.61* | Banks | 1480.89* |
| Gasoline | 89.18* | Entertainment | 155.24* |
| Grocery | 509.71* | Stores and Dealers | 2335.61* |
| Health and Personal | 674.54* | Beauty | 733.07* |
| Medical Facilities | 178.43* | Gyms | 136.75* |
|  |  | Restaurant | 226.42* |

**\*Values are statistically significant at p <= 0.05**

**Secondary clustering of essential and non-essential lifestyles**

The analysis performed a secondary clustering of k-means to distinguish the recovery trajectories. Within each primary cluster, the analysis identified three recovery trajectories, and considering all the clusters, the analysis found four patterns of recovery trajectories: (1) short recovery of essential and non-essential lifestyles (~0-2 weeks), (2) long recovery of essential lifestyles (>15 weeks) and recovery of non-essential lifestyles (2–-4 weeks), (3) long recovery of essential and non-essential lifestyles (>15 weeks), and (4) moderate recovery of essential and non-essential lifestyles (7–-8 weeks). Figure 4 shows the spatial distribution of the primary and secondary clusters, and Figure 5 shows the median values of the recovery of essential and non-essential lifestyles. Table 2 shows the ANOVA testing that determined the difference between the recovery trajectories of the secondary clustering. The difference in the means of the recovery trajectories is significant at $p<0.05$. The results indicate there are distinctive recovery trajectories for secondary clustering which imply disproportionate rates of recovery within the primary lifestyle clusters.

**Table 2.** ANOVA testing on essential and non-essential lifestyle recovery for secondary clusters within primary clusters

| Clusters | Essential F-value | Non-Essential F-value |
| --- | --- | --- |
| Cluster 1 | 4129.18* | 4474.89* |
| Cluster 2 | 80.10* | 41.59* |
| Cluster 3 | 1740.17* | 1350.98* |
| Cluster 4 | 313.39* | 365.59* |

**\*Values are statistically significant at p <= 0.05**

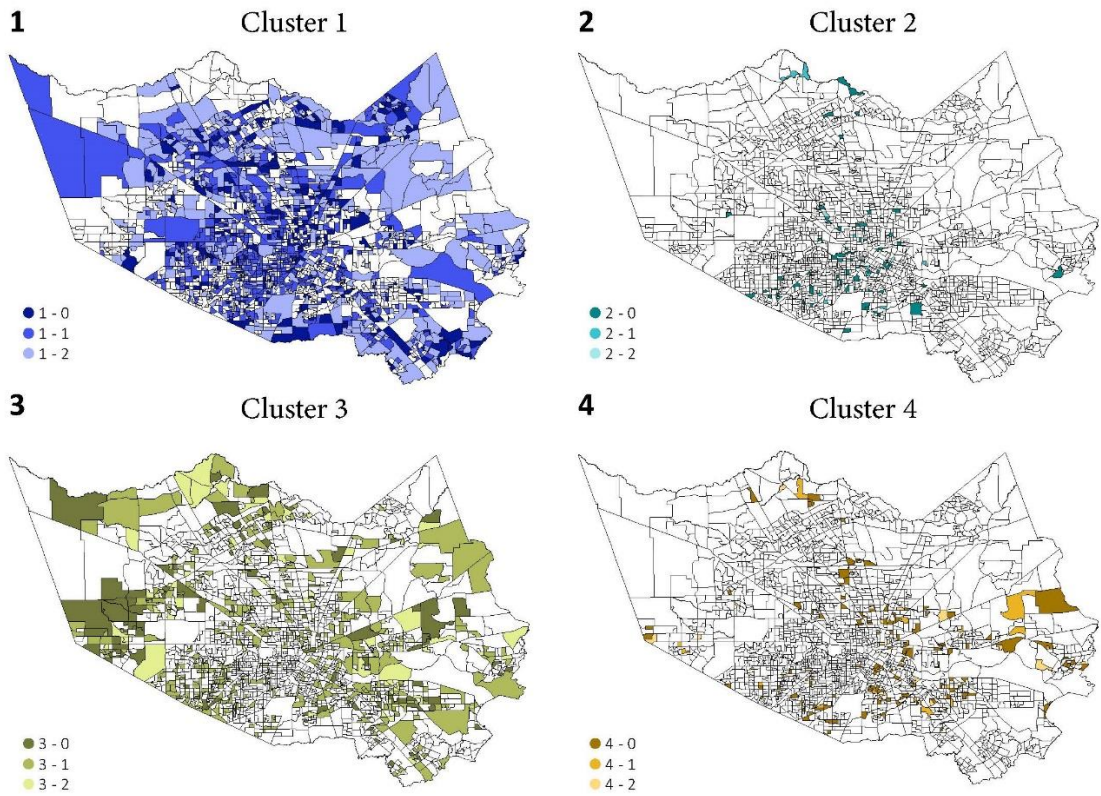

**Figure 4.** Spatial distribution of the secondary clusters
**Notes:** The maps show the spatial distribution of the sub-clusters for each of the four primary clusters. Clusters 1 and 3 represent the majority of primary lifestyle clusters.

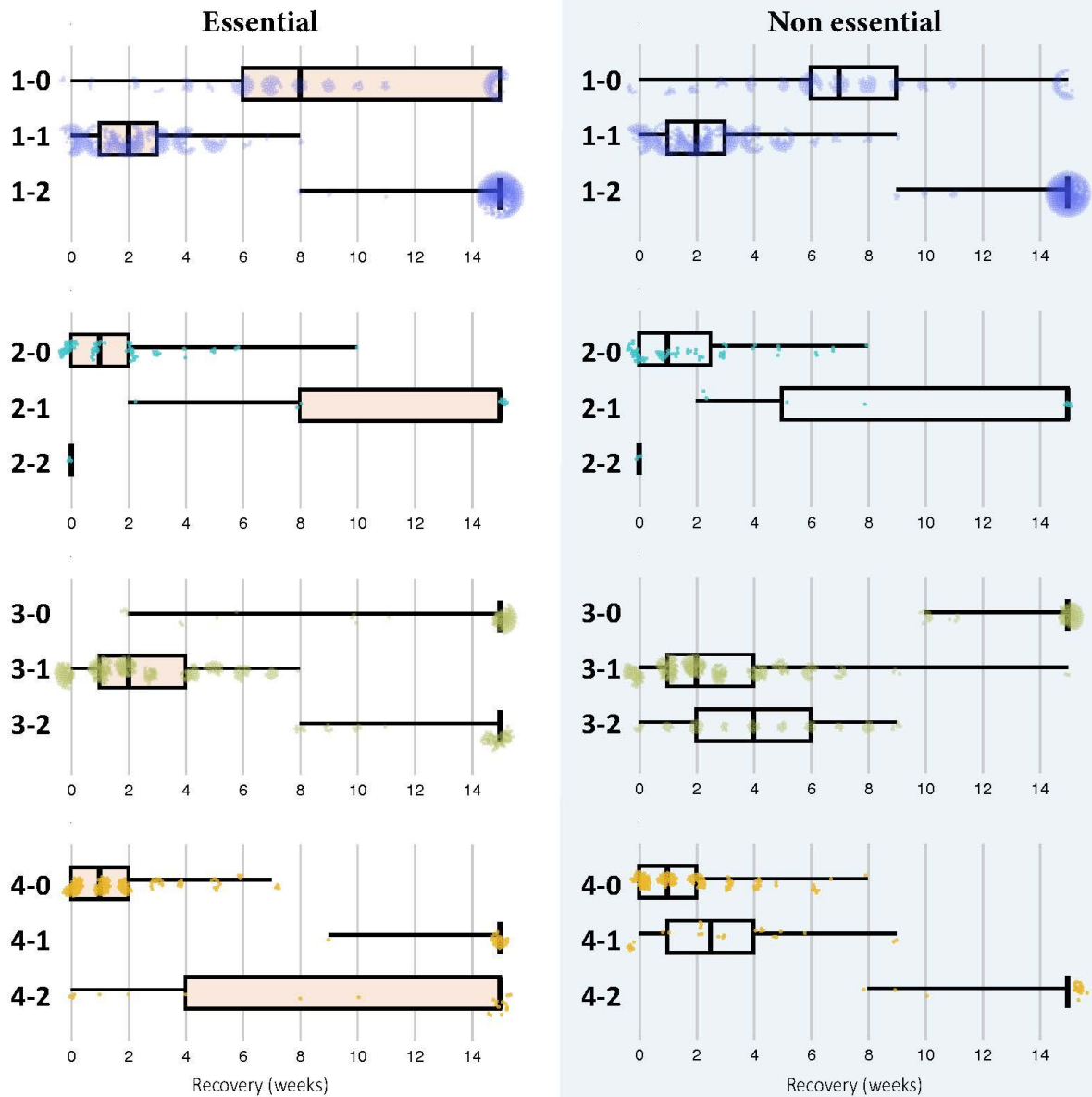

**Figure 5.** Box-whisker-plots of the duration of recovery for secondary clusters
**Notes:** The plots display the distribution of the duration of recovery for essential and non-essential lifestyles of the CBGs in the secondary clusters. The number of CBGs is represented by the cluster of points on the plots.

The research also examined the connection between demographics and lifestyle recovery patterns. Table 3 summarizes the median values of the demographics for the secondary clusters. It also presents the count of CBGs, the median values and percentages of demographics across the CBGs, and the percentage of CBGs in the secondary cluster that have a 1% flooding threshold. The 1% flooding threshold means that at least 1% of the land area of the CBG experienced flooding impacts. The maximum point of disruption is defined as the greatest percentagepercent change in visits to essential and non-essential POIs, while duration of recovery is the number of weeks for a CBG to recover to 90% of its baseline value. In total, essential lifestyles had a median recovery of 5 weeks while non-essential services had a median lifestyle recovery of 4 weeks. Also, in some CBGs, essential lifestyle recovery occurred prior to non-essential lifestyles, vice versa. This result shows that lifestyle recovery may not be as sequential.

Table 3. Median values of the demographics for the secondary clusters

| | Count of CBGs | Total Pop | Elderly (%) | White (%) | Median Income ($) | Flood >=1% (% of CBGs) | Essential: Duration of Recovery | Essential: Maximum Point of Disruption | Non-essential: Duration of Recovery | Non-essential: Maximum Point of Disruption |
|---|---|---|---|---|---|---|---|---|---|---|
| All* | 2144 | 1,843 | 10.01 | 68.21 | 54,539 | 38.66 | 5 | 71.69 | 4 | 85.78 |
| 1-0 | 252 | 1,723 | 10.47 | 69.00 | 56,295 | 44.05 | 8 | 70.97 | 7 | 89.08 |
| 1-1 | 521 | 1,804 | 10.77 | 70.29 | 60,444 | 35.12 | 2 | 65.82 | 2 | 82.41 |
| 1-2 | 519 | 2,040 | 10.03 | 68.26 | 61,655 | 40.85 | 15 | 75.12 | 15 | 89.15 |
| 2-0 | 67 | 1,126 | 10.80 | 57.35 | 43,000 | 38.80 | 1 | 45.76 | 1 | 51.17 |
| 2-1 | 9 | 1,097 | 10.90 | 40.07 | 32,674 | 55.56 | 15 | 68.31 | 15 | 68.62 |
| 2-2 | 3 | 1,453 | 9.93 | 81.76 | 67,857 | 33.33 | 0 | -1.18 | 0 | -4.43 |
| 3-0 | 115 | 2,233 | 9.20 | 68.96 | 61,319 | 42.61 | 15 | 65.08 | 15 | 82.84 |
| 3-1 | 389 | 1,853 | 9.55 | 65.81 | 47,721 | 40.36 | 2 | 54.05 | 2 | 77.62 |
| 3-2 | 95 | 2,093 | 8.77 | 70.55 | 49,821 | 34.74 | 15 | 62.42 | 4 | 82.25 |
| 4-0 | 127 | 1,441 | 9.00 | 62.23 | 38,005 | 29.13 | 1 | 55.08 | 1 | 53.13 |
| 4-1 | 26 | 1,793 | 8.94 | 56.51 | 35,256 | 30.77 | 15 | 63.05 | 2.5 | 70.64 |
| 4-2 | 21 | 1,349 | 11.76 | 61.71 | 37,283 | 33.33 | 15 | 67.03 | 15 | 72.43 |

- Yellow shading inidcates that the secondary cluster had long recovery (>15 weeks) in only their essential services to 90% of the baseline
- *All referring to all CBGs in Harris County

The results showed some variation between the demographic attributes of populations and their lifestyle recovery; however, the results did not suggest a general pattern of lifestyle recovery for different race and income groups to yield statistically conclusive findings. ANOVA testing found no statistical differences between the mean or median values of the demographic characteristics for the three lifestyle recovery trajectories within each lifestyle cluster. However, it is worth noting the demographic differences from an exploratory perspective. For example, Cluster 1-2 and Cluster 3-0 had long recovery periods (>15 weeks), and these clusters had median incomes above the median income of all CBGs. However, Cluster 2-1 and Cluster 4-2 also had long recovery periods (>15 weeks) but these clusters had median incomes below the median incomes of all CBGs. Though these secondary clusters had different ranges of income, they still exhibited the same lifestyle recovery. This result suggests that people with different demographic attributes can have similar lifestyle patterns as well as similar lifestyle recovery.The relationship between demographics and lifestyle recovery can be examined within the same lifestyle patterns, or primary cluster. Cluster 2-1, which had a $32,674 median income and a 60% minority populations, experienced long period of recovery periods (>15 weeks) while Cluster 3-2 had short recovery periods (2 weeks for both essential 2-2, which had a median income of $67,857 and 19% minority population, but a recovery period of 0 weeks. This demonstrates the possible disparities of recovery trajectories for vulnerable groups within similar lifestyle. However, the opposite trend is observed for Cluster 3. Cluster 3-0 , which had a median income of $61,319 and a long period of recovery (>15 weeks)  while Cluster 3-1, which had a median income of $47,721, had a short period of recovery (2 weeks). Indeed, future research could examine the underlying mechanisms of demographics and lifestyle recovery within the same lifestyle patterns.

Furthermore, the results indicate that the flooding attributes of the CBGs are not always associated with lifestyle recovery. For example, Cluster 2-2 had 33% of CBGs with at least 1% flooding and had a recovery period of 0 weeks while Cluster 4-2 which similarly experienced 33% of CBGs with at least 1% flooding had a long recovery period (>15 weeks). In total, approximately 89% of non-flooded CBGs still experienced at least 1 week of disruption, which emphasizes that impacts on lifestyle can extend beyond direct flooding.

We further examined disparities in lifestyle recovery by focusing on the CBGs in secondary clusters which had long periods of recovery (>15 weeks). Approximately 500 CBGs from Cluster 1-1, four2, 4 CBGs from Cluster 2-1, 92 CBGs from Cluster 3-0, and 12 CBGs from Cluster 4-0 did not recover to 90% of their essential and non-essential lifestyles after 15 weeks. Table 4 shows the median values of the demographics from these CBGs as well as the count of CBGs, percentage from the secondary and primary cluster, and the percentage of CBGs with 1% flooding. Figure 6 displays the spatial distribution of the CBGs with long recovery durations (>15 weeks) including their associated secondary clustering.

**Table 4.** Median values of the demographics for CBGs with long recovery (>15 weeks for essential and non-essential services)

|  | Count of CBGs | CBGs from Secondary Cluster (%) | CBGs from Primary Cluster (%) | Total Pop | Elderly (%) | White (%) | Median Income ($) | Flood >=1% (% of CBGs) |
|---|---|---|---|---|---|---|---|---|
| **All*** | 2144 | - | - | 1,843 | 10.01 | 68.21 | 54,539 | 38.66 |
| **1-2** | 500 | 95.97 | 38.70 | 2,054 | 9.97 | 68.08 | 61,674 | 40.60 |
| **2-1** | 4 | 133.33 | 5.06 | 901 | 11.80 | 31.90 | 29,854 | 50.00 |
| **3-0** | 92 | 80.0 | 15.35 | 2,272 | 9.00 | 69.57 | 61,637 | 45.65 |
| **4-2** | 12 | 9.45 | 6.90 | 1,435 | 11.71 | 69.03 | 45,774 | 33.33 |

*All referring to all CBGs in Harris County

Although all these CBGs were greatly impacted, the distinct demographics of these home CBGs could inform how emergency managers and public officials priortize and manage the recovery of these areas as they can throughly understand which populations are being impacted. For example, Cluster 2-2 is not only greatly impacted from the long period of recovery but is also characterized by lower median income and high percentage of minority residents compared to the median of all CBGs in Harris County. CBGs with a long period of recovery may not always be subjected to any direct flooding. In essence, 297 CBGs in Cluster 1-2, 2 CBGs in Cluster 2-1, 50 CBGs in Cluster 3-0, and 8 CBGs in Cluster 4-2, which equates to almost 59% of the CBGs with long recovery periods, did not experience at least 1% flooding of the area. The findings further indicate that the mechanisms driving the recovery of lifestyles is beyond the extent of flooding even in the most extreme circumstances. It demonstrates that the spatial structure of communities, which is primarily formed by human mobility and distribution of POI facilities, extends the spatial reach of flood impacts on people's lifestyle regardless of the flood status of their home CBG.

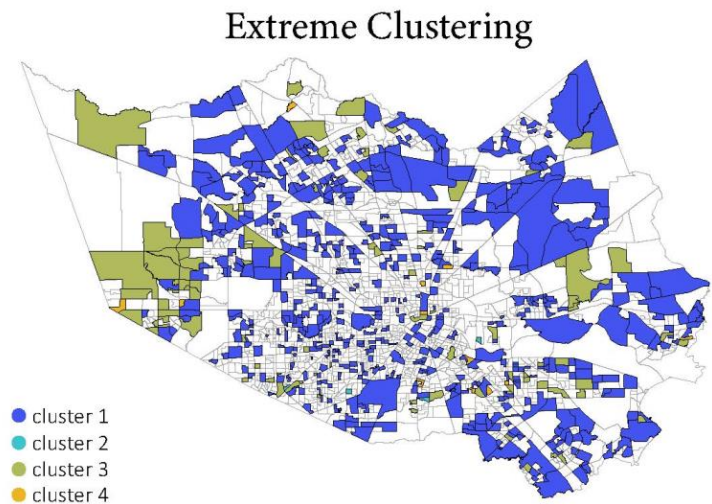

**Figure 6.** Spatial distribution of CBGs with long recovery duration (>15 weeks of essential and non-essential)
**Notes:** The above map shows the spatial distribution of the CBGs with long recovery duration (> 15 weeks) labeled with their associated primary cluster.

## DISCUSSION AND CONCLUDING REMARKS

Examination of community recovery in the aftermath of disasters at the population lifestyle level provides a rather transformative approach to quantification and proactive monitoring of recovery. The research departs from the current approaches which rely solely on qualitative measures of recovery, put the burden of data collection on impacted people, and provide lagging indicators of recovery. In the study presented in this paper, we used fine-grained location intelligence data to capture distinct patterns of population lifestyles in spatial areas (CBGs), and accordingly, specified and quantified the extent of flooding impact and duration of recovery of population lifestyle patterns. The recovery of population lifestyle patterns would be a critical recovery milestone capturing the combined effects of people's activities, restoration of infrastructure, and access to businesses and services in the community. Hence, the study outcomes include multiple methodological, theoretical, and practical contributions.

From the methodological perspective, this study presents a novel methodology to specify and quantify recovery trajectories based on lifestyle analysis as a critical milestone of recovery. We leveraged location intelligence data at the CBG level to specify distinct patterns of lifestyles and to quantify the extent of disaster impacts and speed of recovery. While location intelligence data has been utilized for examining other disaster-related phenomena such as evacuation [19,20,24,44], flooding impacts [18,45-47], community resilience [1,10,23], and access to facilities [12,25,43], no prior work had attempted to leverage location intelligence data in the context of recovery. In particular, the focus on population lifestyle patterns is particularly novel in this study. The methodology presented in this study facilitates the location-based data processing needed to obtain the population visitation patterns and to uncover lifestyle patterns over time and space based on the overlying human mobility, POI facility location, building footprint, and socio-demographic datasets.

Our work also presents multiple theoretical findings of significance. First, the results show that distinct lifestyle patterns of a community are not very diverse, but the level of dependence of different lifestyle clusters on facilities vary. For example, there are four distinctive lifestyles in Harris County as confirmed through statistical analysis. It is important to consider the overall ranking of the essential and non-essential POIs within each lifestyle cluster, as well as the subtle differences in the dependences of populations on POIs within each lifestyle cluster. For example, lifestyle Cluster 2 ranked medical facilities as the most frequently visited essential lifestyle within the cluster (32%); this indicates the high dependency of the populations in CBGs of this cluster to medical facilities. In addition, there were varied relative frequencies of POI visits across different clusters. For example, Cluster 1 has the highest relative frequency of visits to health and personal stores (77%) and stores and dealers (66%) and Cluster 4 has the highest relative frequency of visits to grocery stores (28%) and beauty care (22%) when compared to the other clusters. A disruption in standard lifestyles could negatively impact the health and well-being of populations in these lifestyle clusters because residents have significant dependence on acquiring certain goods and services.

Second, the results revealed diverse and distinct recovery trajectories of population lifestyles. The study signifies lifestyle recovery as a critical milestone in the aftermath of disasters and advances our understanding of differential rates of recovery among locations and populations by analyzing recovery of lifestyles at the CBG level. The results show diverse recovery trajectories within each lifestyle cluster. The analysis revealed four recovery patterns. These included (1) CBGs with short recovery of essential and non-essential lifestyles (~0-2 weeks), (2) CBGs with long recovery of essential lifestyles (lifestyle recovery >15 weeks) and relatively short recovery of non-essential lifestyles (2–4 weeks), (3) CBGs with long recovery of essential and non-essential lifestyles (>15 weeks), and (4) CBGs with moderate recovery duration of essential and non-essential lifestyles (7–8 weeks). These recovery trajectories indicate that the recovery of essential and non-essential lifestyles do not have a fixed sequence (i.e., essential lifestyle recovery occurs prior to non-essential lifestyle recovery). As shown in the recovery pattern results, for some CBGs, even when non-essential lifestyles are recovered, essential lifestyles could be still disrupted. The results also indicates that within the same essential and non-essential lifestyle clusters, disproportionate rates of recovery are observed. In other words, the rate of recovery of lifestyles does not depend on the lifestyle characteristics of populations. CBGs with populations of similar lifestyles could have different lifestyle recovery rates.

Third, the research revealed instances where different demographic attributes had similar lifestyle patterns or similar lifestyle recovery trajectories and durations. Findings related to the correlation between the demographics and duration of recovery were not statistically significant ($p<0.05$), which shows that there may not be a general pattern between demographic attributes and lifestyle recovery. However, it is important to view the data from an exploratory perspective to effectively target specific areas that are especially vulnerable to lifestyle recovery. In particular, the study examined the demographic attributes of extremely impacted CBGs from the secondary clusters which had long recovery duration (>15 weeks). For example, Cluster 1-2 and Cluster 3-0 had below median incomes while Clusters 2-1 and Cluster 4-2 had above median incomes; yet these clusters had long recovery rates (>15 weeks) for their different lifestyle patterns. Examples such as these demonstrate the need for understanding lifestyle patterns in conjuction with demographics, as lifestyles signal the different needs of populations, while demographics are rooted in recovery equity. Disaster research has suggested that socially vulnerable populations, including those of low median income, high percentage of minority residents, and high percentage of elderly residents, as well as locations of extensive flooding, have unique challenges as they recover from the disaster impact [21,48,49]. Delays in lifestyle recovery of vulnerable populations could lead to significant wellbeing impacts. The approaches for restoring lifestyle patterns could be influenced by the demographics attributes of the residents in terms of their unique hardships and capacities. Future studies can further examine the relationship between lifestyle recovery and other community attributes such as socio-demographic and built environment characteristics and infrastructure restorations.

Fourth, the findings of this study revealed that the flood impacts on lifestyles extends beyond direct flood exposure. The spatial structures of communities formed by human mobility and spatial distribution of facilities extends the spatial reach of flood impacts on population lifestyles. Hence, populations living in non-flooded CBGs could experiences considerable disruptions in their essential and non-essential lifestyle activities. The association between flooding and the lifestyle recovery was also not statistically significant. In fact, approximately 89% of non-flooded CBGs still experienced at least 1 week of disruption to the essential and non-essential lifestyles. When examining the CBGs with long recovery periods (>15 weeks), about 58% of these CBGs did not experience at least 1% of flooding impacts. This is because the lifestyles are dynamic in their spatiotemporal characteristics, and thus, even areas that were not directly impacted by flooding could face lifestyle disruption. Residents may be unable to travel and to access essential and non-essential facilities to meet their lifestyle needs. While the majority of literature has focused primarily on impacts and recovery from direct flood exposure, this study reveals the systemic effects of flooding on the broader community. This finding signifies the need for more theoretical understanding of the intersection of hazard exposure and human network dynamics that could extend the spatial reach and spillover effects of hazards such as flooding beyond the direct exposure on communities.

From the practice perspective, this study offers a more data-driven and proactive approach to community recovery in the aftermath of disasters. The current approach to recovery is rather reactive due to the reliance on qualitative measures and dependence on surveys completed impacted residents. This approach puts the burden of data collection on residents; insights gleaned from surveys could provide only lagging indications of recovery. The literature suggests that inefficient and delayed allocation of recovery resources is one of the barriers to timely community recovery. However, using the location-based data and approach used in this study, public officials and emergency managers could proactively monitor the recovery of different spatial areas based on how quickly lifestyle patterns of populations return to the pre-disaster patterns. Using this data-driven and quantitative insights, recovery resources could more efficiently be allocated to areas with slower recovery. Facilitating faster lifestyle recovery could reduce the wellbeing impacts of disasters on communities (and vulnerable populations in particular).

**Data availability**
All data were collected through a CCPA- and GDPR-compliant framework and utilized for research purposes. The data that support the findings of this study are available from Spectus Inc., but restrictions apply to the availability of these data, which were used under license for the current study. The data can be accessed upon request submitted on Spectus.ai. Other data we use in this study are all publicly available.

**Code availability**
The code that supports the findings of this study is available from the corresponding author upon request. The authors also would like to acknowledge the data support from Spectus Inc.

**Acknowledgments**

This material is based in part upon work supported by the National Science Foundation under Grant CMMI-1846069 (CAREER) and the support of the National Science Foundation Graduate Research Fellowship. The authors also would like to acknowledge the data support from Spectus Inc. Any opinions, findings, conclusions, or recommendations expressed in this material are those of the authors and do not necessarily reflect the views of the National Science Foundation, or Spectus Inc.

**Author Contributions:** All authors critically revised the manuscript, gave final approval for publication, and agree to be held accountable for the work performed therein. N.C. was the lead PhD student researcher and first author, who was responsible for human mobility and statistical analysis. C. L and Y. Z provided support in the data processing of the human mobility data and coding assistance. A. M was the faculty advisor for the project.

**Supplementary Information**

Table SI.1 shows the prominent essential and non-essential POIs of the analysis along with the NAICS digits and corresponding name of the category. The study identified 11 lifestyles which contained 5 essential services (gasoline stations, grocery and merchandise, health and personal care stores, medical facilities, and education) and 6 non-essential services (banks, stores and dealers, restaurants, entertainment, recreation and gym centers, and beauty care).

**Table SI.1:** North American Industry Classification System (NAICS) of the essential and non-essential POIs

| Group | NAICS (Last Digits) | Name of the Category according to NAICS |
|---|---|---|
| **Essential Services** | | |
| Gasoline Stations | 4471 | Gasoline Stations |
| Grocery and Merchandise | 4451 | Grocery Stores |
| | 4452 | Specialty Food |
| | 4523 | General Merchandise Stores, including Warehouse Clubs and Supercenters |
| Health and Personal Care Stores | 4461 | Health and Personal Care Stores |
| Medical Facilities | 6212 | Offices of Dentists |
| | 6221 | General Medical and Surgical Hospitals |
| Education | 6113 | Colleges, University, and Professional Schools |
| | 6112 | Junior Colleges |
| **Non-Essential Services** | | |
| Insurance Agencies | 5242 | Agencies, Brokerages, and Other Insurance Related Activities |
| Banks | 5221 | Depository Credit Intermediation |
| Postal Service | 4911 | Postal Service (United States) |
| Stores and Dealers | 441 | Motor Vehicle and Parts Dealers |
| | 442 | Furniture and Home Furnishings Stores |
| | 443 | Electronic and Appliance Stores |
| | 448 | Clothing and Clothing Accessories Stores |
| | 451 | Sporting Goods, Hobby, Musical Instrument, Back Stores |
| | 4522 | Department Stores |
| | 453 | Miscellaneous Store Retailors |
| Restaurants | 7225 | Restaurants and Other Eating Places |
| Entertainment | 7121 | Museums, Historical Sites, and Similar Institutions |
| | 5121 | Motion Picture and Video Industries |
| | 7112 | Performing Arts, Sports, and Related Industries |
| | 7131 | Amusement Parks and Arcades |
| Recreation and Gym Centers | 7139 | Other Amusement and Recreation Industries |
| Beauty Care | 8121 | Personal Care Services |

Figure SI.1 shows the results of the elbow method to determine the optimal k-value for the primary clusters. The y-axis (k) is the number of clusters tested and the x-axis (distortion) is the sum of the square distances from each point to the assigned center. The optimal k-value is chosen when there is a notable "elbow" in the line.

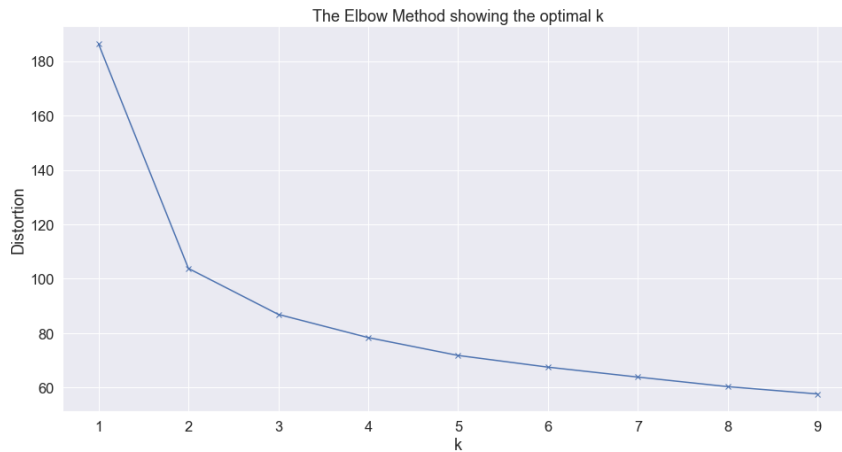

**Figure SI.1:** Elbow method to determine optimal k-value

Figure SI.2 shows the distribution of the demographics for the four primary clusters, and the Table SI.2 is the median values of the demographics.

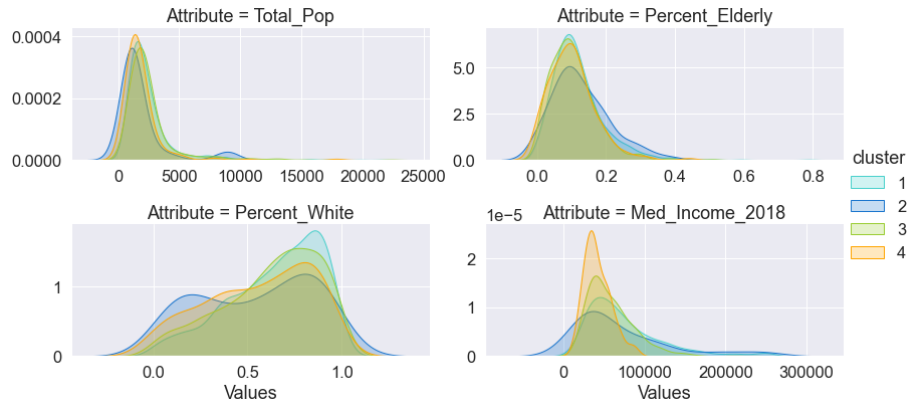

**Figure SI.2:** Distribution of the demographic for primary clusters

**Table SI.2:** Median values of the demographics for the primary clusters

| Cluster | Count of CBGs | Total Population | Elderly (%) | White (%) | Median Income ($) |
|---|---|---|---|---|---|
| 1 | 1,292 | 1,862 | 10.34 | 69.34 | 60,475 |
| 2 | 79 | 1,133 | 10.80 | 57.37 | 43,000 |
| 3 | 599 | 1,973 | 9.36 | 67.09 | 50,500 |
| 4 | 174 | 1,454 | 9.44 | 61.15 | 37,507 |

Tables SI.3a and SI.3b shows the percentage of relative visit frequency to each POI when compared to the total essential POIs or the total non-essential POIs. The red lettering designates that the primary clusters has the highest relative frequency to the specific essential or non-essential POI.

**Table SI.3a:** Percentage of visit relative frequency to essential POIs

| Cluster | Education | Gasoline Stations | Grocery and Merchandise | Health and Personal | Medical Facilities |
|---|---|---|---|---|---|
| 1 | 0.32 | 1.16 | 16.94 | 77.40 | 4.17 |
| 2 | 0.18 | 3.60 | 22.93 | 41.31 | 31.97 |
| 3 | 0.15 | 1.77 | 22.80 | 69.65 | 5.63 |
| 4 | 0.09 | 2.45 | 28.40 | 64.72 | 4.33 |

**Table SI.3b:** Percentage of visit relative frequency to non-essential POIs

| Cluster | Banks | Beauty Care | Entertainment | Recreation and Gyms | Restaurant | Stores and Dealers |
|---|---|---|---|---|---|---|
| 1 | 2.75 | 6.38 | 0.43 | 0.79 | 24.07 | 65.57 |
| 2 | 9.53 | 14.13 | 0.99 | 2.21 | 34.27 | 38.84 |
| 3 | 5.25 | 11.75 | 0.36 | 0.79 | 24.21 | 57.63 |
| 4 | 13.37 | 21.86 | 0.30 | 1.11 | 23.46 | 39.89 |

Figure SI.3 shows the distribution of demographics and flooding for the secondary clusters within the primary clusters.

**Cluster 1:**

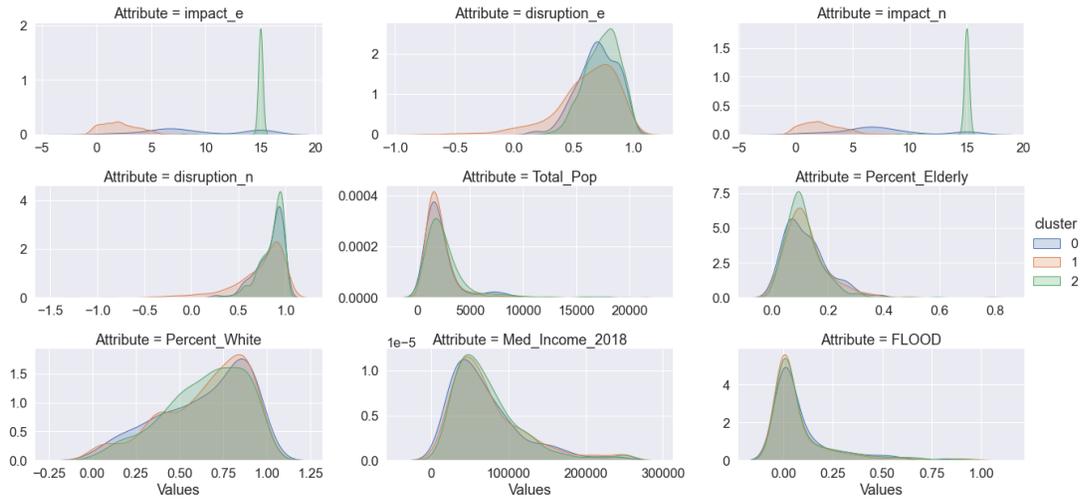

**Cluster 2:**

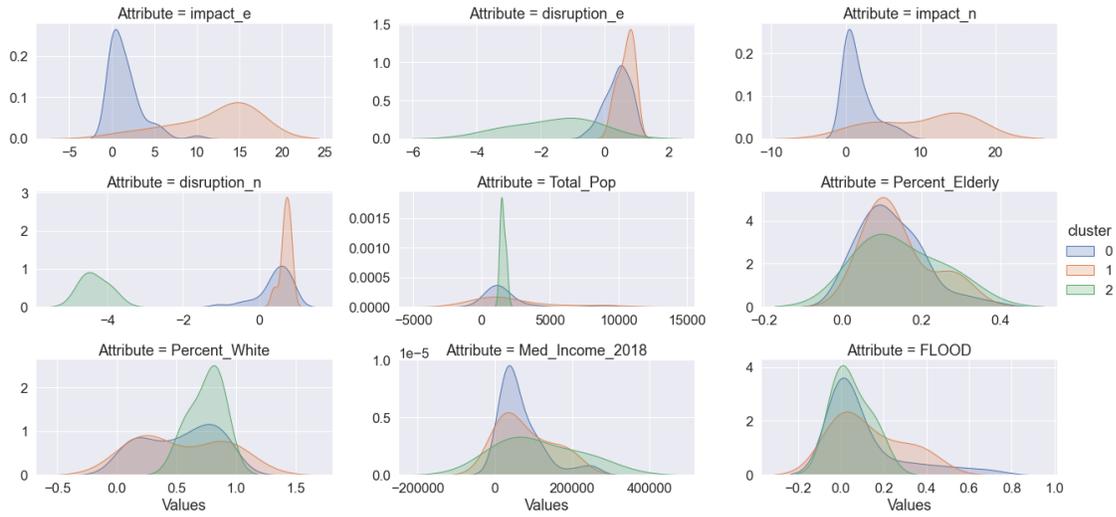

**Cluster 3:**

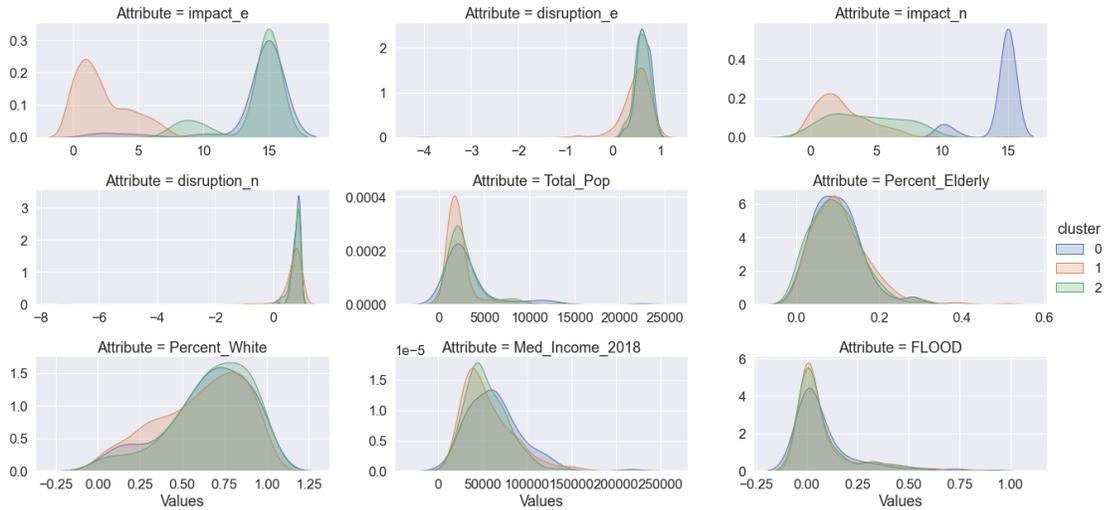

**Cluster 4:**

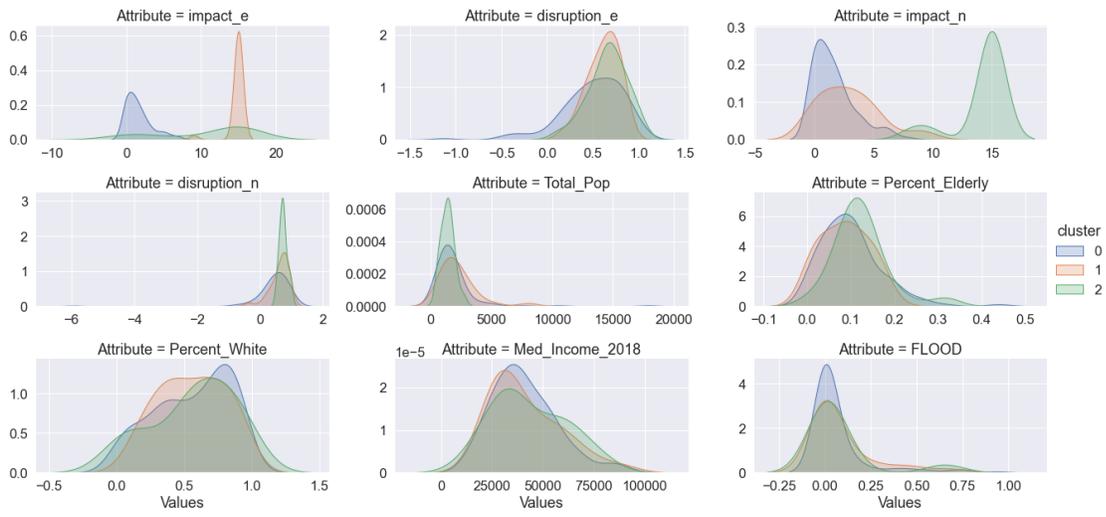

**Figure SI.3:** Distribution of demographics and flooding data for the secondary clusters